\documentclass[letter,10pt,conference]{IEEEtran}

 \usepackage[bookmarks=false]{hyperref}
\ifCLASSINFOpdf
\else
\fi
\usepackage{relsize}
 
\usepackage{centernot}
\usepackage{cite}
\usepackage{amsfonts}
\usepackage{mathrsfs}
\usepackage{bbm}
\usepackage{pifont}
\usepackage{amsfonts}
\usepackage{amsmath, amsthm}
\usepackage[top=.75in,bottom=.89in,left=.625in,right=.625in]{geometry}
\usepackage{tabularx}
\usepackage{listings}
\usepackage{graphicx}
\usepackage{float}
\usepackage{amssymb}
\usepackage{array}
\usepackage{fancyhdr}
\usepackage{cases}
 \usepackage{supertabular}
\usepackage{url}
\usepackage{fancyhdr}

\usepackage{psfrag}
\usepackage{color}
\usepackage{bbding}
\newcommand{\bcap} {\hspace{2pt} \mathlarger{\cap}
\hspace{2pt}}
 
\newcommand{\given} {{\mathlarger{\boldsymbol\mid}}\hspace{1pt}}
\newcommand{\bgiven} {{\mathlarger{\mathlarger{\mathlarger{\mathlarger{\mid}}}}}}

\newcommand{\bcup} {\hspace{2pt} \mathlarger{\cup}
\hspace{2pt}}

\newcommand {\C} {{\rm I\kern-5.5pt C}}

\def\centerhack#1{\hbox to 0pt{\hss\footnotesize #1\hss}}
\def\centerhackn#1{\hbox to 0pt{\hss #1\hss}}
\def\dchack#1{\vbox to 0pt{\vss{\hbox to 0pt{\hss#1\hss}}\vss}}

\newtheorem{lem}{Lemma}
\newtheorem{thm}{Theorem}

\newtheorem{rem}{Remark}
\newtheorem{cor}{Corollary}

\newtheorem*{proposition1.1}{Proposition 1.1}
\newtheorem*{proposition1.2}{Proposition 1.2}
\newtheorem*{proposition1.3}{Proposition 1.3}
\newtheorem*{proposition2.1}{Proposition 2.1}
\newtheorem*{proposition2.2}{Proposition 2.2}
\renewcommand\baselinestretch{.98}

\begin{document}

%


\title{$k$-Connectivity of Random Key Graphs}






 \author{ \IEEEauthorblockN{Jun Zhao}
\IEEEauthorblockA{CyLab and Dept.
of ECE \\
Carnegie Mellon University \\
{\tt junzhao@cmu.edu}} \and \IEEEauthorblockN{Osman Ya\u{g}an}
\IEEEauthorblockA{CyLab and Dept.
of ECE\\
Carnegie Mellon University \\
{\tt oyagan@ece.cmu.edu}} \and \IEEEauthorblockN{Virgil Gligor}
\IEEEauthorblockA{CyLab and Dept.
of ECE \\
Carnegie Mellon University \\
{\tt gligor@cmu.edu}}}

\maketitle \thispagestyle{plain} \pagestyle{plain}

%
%



\maketitle


 \begin{abstract}
 
Random key graphs represent topologies of secure wireless sensor networks that
apply the seminal Eschenauer--Gligor
random key predistribution scheme to secure communication between sensors. These graphs have received much attention and
also been used in diverse application areas beyond secure sensor networks; e.g., cryptanalysis, social networks, and recommender systems. Formally, a random key graph with $n$ nodes is constructed by assigning each node $X_n$ keys selected uniformly at random from a pool of $Y_n$ keys and then putting an undirected edge between any two nodes sharing at least one key. Considerable progress has been made in the literature to analyze connectivity and
$k$-connectivity of random key graphs; e.g., Ya\u{g}an and Makowski [ISIT '09, Trans. IT'12] on connectivity under $X_n = \Omega(\sqrt{\ln n}\hspace{1.5pt})$, Rybarczyk [Discrete Mathematics '11] on connectivity under $X_n \geq 2$, and our recent work [CDC '14] on $k$-connectivity under $X_n = \Omega(\sqrt{\ln n}\hspace{1.5pt})$, where $k$-connectivity of a graph ensures connectivity even after the removal of $k$ nodes or $k$ edges. Yet, it still remains an open question for $k$-connectivity in random key graphs under $X_n \geq 2$ and $X_n = o(\sqrt{\ln n}\hspace{1.5pt})$ (the case of $X_n=1$ is trivial). In this paper, we answer the above problem by providing an exact analysis of $k$-connectivity in random key graphs under $X_n \geq 2$.

\end{abstract}
 


\section{Introduction} \label{introduction}

Random key graphs, also known as homogeneous random intersection
graphs, have been investigated widely in the literature
\cite{r1,ryb3,zz,yagan_onoff,yagan,ZhaoYaganGligor,ISIT,r10,r4}.
The notion of random key graph results from the seminal
Eschenauer--Gligor (EG) random key predistribution scheme \cite{virgil},
which is the most recognized solution to secure communication using cryptographic keys in wireless sensor networks  \cite{yagan}.
The definition of a random key graph can also be generalized beyond cryptographic keys. Consider 
 a random key graph
$G(n, X_n, Y_n)$ that is constructed on a set of $n$ nodes as follows.
Each node is independently assigned a set of $X_n$ distinct objects,
selected {\em uniformly at random} from a pool of $Y_n$ objects,
where $X_n$ and $Y_n$ are both functions of $n$. An undirected edge exists
between two nodes if and only if they possess at least one common
object. An object is a cryptographic key in the application of random key graphs to
the Eschenauer--Gligor random key predistribution scheme. In addition to the area of secure sensor networks,
random key graphs have 
also been used in various applications including cryptanalysis \cite{r10}, social networks \cite{ZhaoYaganGligor}, and recommender systems \cite{r4}.

($k$-)Connectivity of a random key graph has received much interest
\cite{r1,ryb3,zz,yagan_onoff,yagan,ZhaoYaganGligor,ISIT}.
 A graph is said to be $k$-connected if it remains connected despite
the deletion of at most $(k-1)$ nodes or
edges\footnote{$k$-connectivity given here is equivalent to $k$-{\em
vertex}-connectivity, which can also be defined when \emph{only} node
failure is considered; i.e., the ability of the graph remaining
connected in spite of the removal of at most $(k-1)$ nodes. $k$-{\em
edge}-connectivity is defined similarly for graphs that are still
connected despite the failure of any $(k-1)$ edges.
 It is plain to prove that $k$-vertex-connectivity implies $k$-edge-connectivity
\cite{erdos61conn}.}; an equivalent definition of $k$-connectivity
is that for each pair of nodes there exist at least $k$ mutually
disjoint paths connecting them \cite{erdos61conn}. In the case of
$k$ being 1, $k$-connectivity becomes connectivity, meaning that
each node in the graph can find at least one path to any other node,
either directly or with the help of other relaying nodes. A graph
property related to and implied by $k$-connectivity
 is
that the minimum degree of the the graph is at least $k$ (i.e.,
 each node is directly connected to no less than $k$ other
nodes), where the minimum degree refers to the minimum among the
numbers of neighbors that nodes have.

We investigate $k$-connectivity of random key graphs. Our
contribution is, for a random key graph, to derive the
asymptotically exact probabilities for $k$-connectivity and the
property that the {minimum degree} is at least $k$.

The rest of the paper is organized as follows. Section
\ref{sec:main:res} presents the results. We elaborate the proof of
Theorem \ref{THM_RKGk} in Section \ref{sec_est}. Section
\ref{sec:expe} provides numerical findings to support the
theoretical results. Section \ref{related} surveys related work; and
Section \ref{sec:Conclusion} concludes the paper.

 %
%
%
%


\section{The Results} \label{sec:main:res}

For a random key graph $G(n,X_n,Y_n)$, Theorem \ref{THM_RKGk} and
Corollary \ref{COR_RKGk} below present the asymptotically exact
probabilities for $k$-connectivity and the
property of the {minimum degree} being at least $k$, where $k$ is a
positive integer and does not scale with $n$. The term $\ln$ stands for
the natural logarithm function, and $e$ is its base. We use the standard asymptotic notation $O(\cdot), o(\cdot), \Omega(\cdot),
\omega(\cdot), \Theta(\cdot), \sim$; in particular, for two
positive sequences $x_n$ and $y_n$, the relation $x_n \sim y_n$
means $\lim_{n \to
  \infty} (x_n/y_n)=1$. All asymptotic statements are understood with $n \to \infty$. Also, $\mathbb{P}[\mathcal {E}]$
denotes the probability that event $\mathcal {E}$ occurs.

\begin{thm}\label{THM_RKGk} For a random key graph $G(n,X_n,Y_n)$,
let $q_n$ be the probability that there exists an edge between two
nodes. With a sequence $\alpha_n$ defined by
\begin{align}
q_n &
 = \frac{\ln  n + {(k-1)} \ln \ln n + {\alpha_n}}{n},
\label{thm_eq_ps}
\end{align}
then under 
$ X_n \geq 2$, it follows that
\begin{align} 
 & \lim_{n \to \infty}\mathbb{P} \big[\hspace{2pt}G(n,X_n,Y_n)\text{
is $k$-connected}.\hspace{2pt}\big] \nonumber \\ 
 &=  \lim_{n \to \infty}  \mathbb{P}
\left[\hspace{2pt}G(n,X_n,Y_n)\text{ has a minimum degree at least
}k.\hspace{2pt}\right] 
 \nonumber \\ 
 &
\quad =
\begin{cases} 0, &\text{ if $\lim_{n \to \infty}{\alpha_n}
=-\infty$}, \\  1, &\text{ if $\lim_{n \to \infty}{\alpha_n}
=\infty$,}  \\ e^{- \frac{e^{-\alpha ^*}}{(k-1)!}},
 &\text{ if $\lim_{n \to \infty}{\alpha_n}
=\alpha ^* \in (-\infty, \infty)$.} \end{cases} 
\label{doublexp}
 \end{align}
\end{thm}

We have the following corollary by replacing the condition (\ref{thm_eq_ps}) on the edge probability $q_n$ with a condition on  \vspace{1.7pt} the asymptotics $\frac{{X_n}^2}{Y_n} $ of $q_n$ (formally, $q_n \sim \frac{{X_n}^2}{Y_n}$ holds \vspace{1pt} under $\frac{{X_n}^2}{Y_n} = o(1)$; see  \cite[Lemma 8]{ZhaoYaganGligor}.)

\begin{cor}\label{COR_RKGk} For a random key graph $G(n,X_n,Y_n)$,
with a sequence $\beta_n$ defined by
\begin{align}
\frac{{X_n}^2}{Y_n} &
 = \frac{\ln  n + {(k-1)} \ln \ln n + {\beta_n}}{n},
\label{thm_eq_psXnPn}
\end{align}
then under 
$ X_n \geq 2$, it follows that
\begin{align} 
 & \lim_{n \to \infty}\mathbb{P} \big[\hspace{2pt}G(n,X_n,Y_n)\text{
is $k$-connected}.\hspace{2pt}\big] \nonumber \\ 
 & = \lim_{n \to \infty}  \mathbb{P}
\left[\hspace{2pt}G(n,X_n,Y_n)\text{ has a minimum degree at least
}k.\hspace{2pt}\right] 
 \nonumber \\ 
 &
\quad =
\begin{cases} 0, &\text{ if $\lim_{n \to \infty}{\beta_n}
=-\infty$}, \\  1, &\text{ if $\lim_{n \to \infty}{\beta_n}
=\infty$,}  \\ e^{- \frac{e^{-\beta ^*}}{(k-1)!}},
 &\text{ if $\lim_{n \to \infty}{\beta_n}
=\beta ^* \in (-\infty, \infty)$.} \end{cases} 
\nonumber
 \end{align} 
\end{cor}

\begin{rem} \label{rem1}
From Lemma \ref{lem-only-prove-lnlnn-1} (resp., Lemma \ref{lem-only-prove-lnlnn-2}) in the Appendix, we can introduce an
extra condition $\alpha_n = \pm O(\ln \ln n) = \pm o(\ln n)$ (resp., $\beta_n = \pm O(\ln \ln n) = \pm
o(\ln n)$) in proving Theorem
\ref{THM_RKGk} (resp., Corollary \ref{COR_RKGk}).
 \end{rem}
 
 \begin{rem} 
In Theorem
\ref{THM_RKGk} and Corollary \ref{COR_RKGk}, since the results  are in the asymptotic sense, 
 the conditions only need to hold for all $n$ sufficiently
large.
 \end{rem}

Establishing Corollary \ref{COR_RKGk} given Theorem
\ref{THM_RKGk} is straightforward and is given in the Appendix. Below we explain how to obtain Theorem
\ref{THM_RKGk}. Since a necessary condition for a graph to be $k$-connected is that the minimum degree is at least $k$, the proof of Theorem \ref{THM_RKGk} will be completed
once we have the following two lemmas. Lemma  \ref{lemma-1} is from our prior work \cite{ZhaoYaganGligor}. Lemma \ref{lemma-2} simply reproduces the result on the minimum degree in Theorem \ref{THM_RKGk}.

\begin{lem}[\hspace{0pt}{Our work \cite[Lemma 5]{ZhaoYaganGligor}}\hspace{0pt}]  \label{lemma-1}
For a random key graph $G(n,X_n,Y_n)$ under (\ref{thm_eq_ps}) and $ X_n \geq 2$, it follows that
\begin{align}
 \hspace{-2pt}\lim_{n \to \infty} \hspace{-2pt} \mathbb{P}\bigg[\hspace{-4pt}\begin{array}{l}G(n,X_n,Y_n)\text{ has a minimum degree at least
}k,\\\text{but is not $k$-connected}.\end{array}\hspace{-4pt}\bigg] & \hspace{-2pt}=\hspace{-2pt} 0.
  \end{align}
\end{lem}


\begin{lem}  \label{lemma-2}
For a random key graph $G(n,X_n,Y_n)$ under (\ref{thm_eq_ps}) and $ X_n \geq 2$, it follows that
\begin{align}
&  \lim_{n \to \infty}  \mathbb{P}
\left[\hspace{2pt}G(n,X_n,Y_n)\text{ has a minimum degree at least
}k.\hspace{2pt}\right]   \nonumber \\
&   \quad  = \text{right hand side of (\ref{doublexp})}. \nonumber 
 \end{align}
\end{lem}

 
  By \cite[Lemma 2]{mobihocQ1}, Lemma \ref{lemma-2} will follow once we show Lemma  \ref{lemma-3} below, where we let $\mathcal {V} = \{v_1, v_2,
\ldots, v_n \}$ be the set of nodes in a random key graph $G(n, X_n, Y_n)$.
\begin{lem}  \label{lemma-3}
For a random key graph $G(n,X_n,Y_n)$ under (\ref{thm_eq_ps}) and $ X_n \geq 2$, it follows for integers $m\geq 1$ and $h \geq 0$ that
\begin{align}
&  \mathbb{P} [\text{Nodes }v_{1}, v_{2}, \ldots, v_{m}\text{ have
degree }h] \nonumber \\
&   \quad \sim (h!)^{-m}  (n q_n)^{hm} e^{-m n q_n}.
\label{eqn_node_v12n}
\end{align}
\end{lem}

We detail the proof of Lemma  \ref{lemma-3} in the next section.

\section{The Proof of Lemma  \ref{lemma-3}} \label{sec_est}

In a random key graph $G(n, X_n, Y_n)$, recalling that 
 $\mathcal {V} = \{v_1, v_2,
\ldots, v_n \}$ is the set of nodes, we let $S_i$ be the set of $X_n$
distinct objects assigned to node $v_i \in \mathcal {V}$.
We further define $\mathcal {V}_m$ as $ \{v_1, v_2, \ldots, v_m\}$ and
$\overline{\mathcal {V}_m} $ as $ \mathcal {V} \setminus \mathcal {V}_m
$. Among nodes in $\overline{\mathcal {V}_m}$, we denote by $N_i$
the set of nodes neighboring to $v_i$ for $i=1,2,\ldots,m$. We
denote $N_i \bcap N_j$ by $N_{ij}$, and $S_i \bcap S_j$ by $S_{ij}$.

We have the following two observations:
\begin{itemize} 
  \item [i)] If node $v_i$ has degree $h$,
then $|N_{i}| \leq h$, where the equal sign holds if and only if
$v_i$ is directly connected to none of nodes in $V_m \setminus
\{v_i\}$; i.e., if and only if event $\bigcap_{j \in \{1,2,\ldots,m\}\setminus\{i\}}
(S_{ij}=\emptyset)$ happens.  \vspace{2pt}
    \item [ii)] If $|N_{i}| \leq h$ for any $i=1,2,\ldots,m$, then
\begin{align}
& \bigg|\bigcup_{1\leq i \leq m} N_{i}\bigg| \leq \sum_{1\leq i \leq
m}N_{i}  \leq  hm , \label{Nileq}
\end{align}
where the two equal signs in (\ref{Nileq}) \emph{both} hold if and only if \vspace{-1pt}
\begin{align}
\bigg(\bigcap_{1\leq i <j \leq m} (N_{ij}=\emptyset)\bigg)  \bcap
\bigg(\bigcap_{1\leq i \leq m}(|N_{i}| = h)\bigg).  \vspace{-1pt}\label{Nij}
\end{align}
\end{itemize}

From i) and ii) above, if nodes $v_{1}, v_{2}, \ldots, v_{m}$ have degree $h$,
we have either of the following two cases:
\begin{itemize} 
  \item [(a)] Any two of $v_{1}, v_{2}, \ldots, v_{m}$ have no edge in between (namely, $\bigcap_{1\leq i <j \leq m}
(S_{ij}=\emptyset)$); and event (\ref{Nij}) happens. \vspace{1pt}
  \item [(b)] $\big|\bigcup_{1\leq i \leq m} N_{i}\big|
  \leq hm -1$.
\end{itemize}
In addition, if case (a) happens, then nodes $v_{1}, v_{2}, \ldots,
v_{m}$ have degree $h$. However, if case (b) occurs, there is no
such conclusion. With $P_a$ (resp., $P_b$) denoting the probability
of case (a) (resp., case (b)), we obtain  \vspace{-1pt}
\begin{align}
 & P_a \leq \mathbb{P} [\text{Nodes }v_{1}, v_{2}, \ldots, v_{m}\text{ have
degree }h] \leq P_a + P_b,  \vspace{-1pt}\nonumber
\end{align}
where  \vspace{-1pt}
\begin{align}
P_a =  \mathbb{P}\bigg[ \bigg(\bigcap_{1\leq i <j \leq m}
(S_{ij}=\emptyset)\bigg)  &   \bcap \bigg(\bigcap_{1\leq i <j \leq m}
(N_{ij}=\emptyset)\bigg)   \nonumber \\
&    \bcap \bigg(\bigcap_{1\leq i \leq
m}(|N_{i}| = h)\bigg)\bigg],  \vspace{-1pt} \nonumber
\end{align}
and  \vspace{-1pt}
\begin{align}
 P_b & = \mathbb{P}\bigg[\bigg|\bigcup_{1\leq i \leq m} N_{i}\bigg|
  \leq hm -1\bigg].  \vspace{-1pt} \nonumber
\end{align}
Hence, (\ref{eqn_node_v12n}) holds after we prove the
following (\ref{prop2}) and (\ref{prop1}):
\begin{align}
 P_b & =   o \left((nq_n)^{hm} e^{-m n q_n}\right). \label{prop2}
\end{align}
and
\begin{align}
P_a & \sim (h!)^{-m} (n q_n)^{hm} e^{-m n q_n} \cdot
 [1+o(1)], \label{prop1}
\end{align}

We will prove (\ref{prop2}) and (\ref{prop1}) below. 
 We let $\mathbb{S}_m$ denote the tuple $(S_1,S_2,\ldots,S_m)$. The
expression ``$\given \mathbb{S}_m = \mathbb{S}_m^*$'' means ``given
$S_1=S_1^*,S_2=S_2^*,\ldots,S_m=S_m^*$'', where $\mathbb{S}_m^* =
(S_1^*,S_2^*,\ldots,S_m^*)$ with $S_1^*,S_2^*,\ldots,S_m^*$ being
arbitrary $X_n$-size subsets of the object pool. Note that
$S_{ij}^{*} : = S_{i}^{*} \cap S_{j}^{*}$. For two different nodes $v$ and $w$ in the graph $G(n, X_n, Y_n)$, we use $v\leftrightarrow w$ to denote
the event that there is an edge between $v$ and $w$; i.e., the symbol ``$\leftrightarrow $'' means 
``is directly connected to''.

\subsection{The Proof of (\ref{prop2})}

Let $w$ be an arbitrary node in $\overline{V_m}$. We have
\begin{align}
  &  \mathbb{P}\bigg[\bigg|
\bigcup_{1\leq i \leq m} N_{i}\bigg| =  t \bgiven
\mathbb{S}_m= \mathbb{S}_m^*\bigg]  \label{t} \\
& = \frac{(n-m)!}{t!(n-m-t)!}   \nonumber \\
& \times \big\{\mathbb{P}[w\leftrightarrow \text{ at least one of nodes in }V_m \given
\mathbb{S}_m= \mathbb{S}_m^*]\big\}^t  \nonumber \\
&   \times \big\{\mathbb{P}[w\leftrightarrow \text{none of
nodes in }V_m \given \mathbb{S}_m= \mathbb{S}_m^*]\big\}^{n-m-t}.
\label{x}
\end{align}

By the union bound, it holds that
\begin{align}
& \mathbb{P}[w\leftrightarrow \text{at least one of
nodes in }V_m \given \mathbb{S}_m= \mathbb{S}_m^*]  \nonumber \\
& \leq \sum_{1\leq i \leq m}\mathbb{P}
  [w\leftrightarrow v_i \given \mathbb{S}_m= \mathbb{S}_m^*] = m q_n,\label{lll}
\end{align}
which yields
\begin{align}
& \mathbb{P}[w\leftrightarrow \text{none of nodes in }V_m \given
\mathbb{S}_m= \mathbb{S}_m^*] \geq  1 - m q_n. \label{sstar3}
\end{align}
In addition,
\begin{align}
& \mathbb{P}[w\leftrightarrow \text{none of nodes in }V_m \given
\mathbb{S}_m= \mathbb{S}_m^*] \nonumber \\ &  = \frac{\binom{Y_n -
|\bigcup_{1\leq i \leq m} S_i^*|}{X_n}}{\binom{Y_n}{X_n}} \nonumber \\
& \leq (1-q_n)^{{X_n}^{-1}{|\bigcup_{1\leq i \leq m} S_i^*|}} \quad
\text{(by \cite[Lemma 5.1]{yagan_onoff})}  \nonumber \\
& \leq e^{-{X_n}^{-1}q_n{|\bigcup_{1\leq i \leq m} S_i^*|}} \quad
\text{(by $1+x \leq e^x$ for any real $x$)}.
\label{sstar4}
\end{align}

We will prove
\begin{align}
& \sum_{\mathbb{S}_m^*} \Big\{ \mathbb{P}[\mathbb{S}_m =
\mathbb{S}_m^*] \nonumber \\ & \quad \times   \big\{ \mathbb{P}[w\leftrightarrow \text{none of nodes in }V_m \given \mathbb{S}_m= \mathbb{S}_m^*]
\big\}^{n-m-hm}
 \Big\} \label{sstar2} \\
  & \quad \leq  e^{-m
n q_n} \cdot
 [1+o(1)] . \label{sstar}
\end{align}
From (\ref{x}) (\ref{lll}) and (\ref{sstar}), we derive
\begin{align}
P_b  & = \mathbb{P}\bigg[\bigg|\bigcup_{1\leq i \leq m} N_{i}\bigg|
  \leq hm -1\bigg]   \nonumber \\ &= \sum_{t=0}^{hm -1}
   \sum_{\mathbb{S}_m^*} \Big\{ \mathbb{P}[\mathbb{S}_m = \mathbb{S}_m^*]
    \cdot (\ref{t})
 \Big\}
  \nonumber \\
  & \leq \sum_{t=0}^{hm -1} \Big[ n^t \cdot (m q_n)^t  \cdot  (\ref{sstar2})\Big]
 \nonumber \\ &   \leq (nq_n)^{hm} e^{-m n q_n} \cdot
 [1+o(1)] \cdot m^{hm}
  \sum_{t=0}^{hm -1} (mnq_n)^{t-hm}
 . \label{bp}
\end{align}

As noted in Remark \ref{rem1}, we can introduce an extra condition
$\alpha_n = \pm O(\ln \ln n)= \pm o(\ln n)$ in establishing Theorem \ref{THM_RKGk}.
From $\alpha_n = \pm o(\ln n)$ and (\ref{thm_eq_ps}), we obtain
\begin{align}
q_n & \sim  \frac{\ln n}{n}.\label{eq_pe_lnnn}
\end{align}
Applying (\ref{eq_pe_lnnn}) to (\ref{bp}), we obtain (\ref{prop2}). Hence,
we complete the proof of (\ref{prop2}) once showing (\ref{sstar}),
whose proof is detailed below.

From (\ref{sstar3}) (\ref{sstar4}) and (\ref{eq_pe_lnnn}), we have
\begin{align}
(\ref{sstar2}) \hspace{-1pt}& \hspace{-1pt}\leq\hspace{-1pt} (1 - m q_n)^{-m-hm}  \nonumber  \\
  \hspace{-1pt}& \hspace{-1pt}\quad \times \sum_{\mathbb{S}_m^*}
\Big\{ \mathbb{P}[\mathbb{S}_m = \mathbb{S}_m^*] \cdot
e^{-{X_n}^{-1}nq_n{|\bigcup_{1\leq i \leq m} S_i^*|}}
\Big\} \nonumber  \\
  \hspace{-1pt}& \hspace{-1pt}\leq\hspace{-1pt}[1+o(1)]\hspace{-1pt}\cdot\hspace{-1pt}\sum_{\mathbb{S}_m^*}
\Big\{ \mathbb{P}[\mathbb{S}_m \hspace{-1pt}=\hspace{-1pt} \mathbb{S}_m^*] \hspace{-1pt}\cdot\hspace{-1pt}
e^{-{X_n}^{-1}nq_n{|\bigcup_{1\leq i \leq m} S_i^*|}} \Big\},
\end{align}
so (\ref{sstar}) holds once we demonstrate
\begin{align}
& \sum_{\mathbb{S}_m^*} \Big\{ \mathbb{P}[\mathbb{S}_m =
\mathbb{S}_m^*]  \cdot e^{-{X_n}^{-1}nq_n{|\bigcup_{1\leq i \leq m}
S_i^*|}} \Big\}  \nonumber  \\
  & \quad \leq e^{-m n q_n} \cdot
 [1+o(1)] . \label{ms}
\end{align}
We denote the left hand side of (\ref{ms}) by $Z_{m,n}$. Dividing $\mathbb{S}_{m}^*$ into two parts $\mathbb{S}_{m-1}^*$
and $S_m^*$,
 we derive
\begin{align}
  Z_{m,n}
  &=  \sum_{\begin{subarray}{c}\mathbb{S}_{m-1}^*
  \\S_m^*\end{subarray}} \Big\{ \mathbb{P}[(\mathbb{S}_{m-1} = \mathbb{S}_{m-1}^*)
  \bcap(S_m = S_m^*)]  \nonumber  \\
  & \quad\quad\quad\quad \times e^{-{X_n}^{-1}nq_n{|\bigcup_{1\leq i \leq m}
S_i^*|}} \Big\}\nonumber \\  &= \sum_{\mathbb{S}_{m-1}^*}
\mathbb{P}[\mathbb{S}_{m-1} = \mathbb{S}_{m-1}^*] \bigg\{
e^{-{X_n}^{-1}nq_n{|\bigcup_{1\leq i \leq m-1} S_i^*|}} \nonumber \\
&  \quad\quad \times \sum_{S_m^* }
\mathbb{P}[ S_m = S_m^* ] e^{-{X_n}^{-1}nq_n{|S_m^* \setminus
\bigcup_{1\leq i \leq m-1} S_i^*|}}\bigg\} ,\label{HnmHnm1}
\end{align}
 where
\begin{align}
&  \sum_{S_m^* } \mathbb{P}[ S_m = S_m^* ] e^{-{X_n}^{-1}nq_n{|S_m^*
\setminus \bigcup_{1\leq i \leq m-1} S_i^*|}} \nonumber \\ &  \leq
e^{-n q_n}\sum_{S_m^* } \mathbb{P}[ S_m = S_m^* ] e^{{X_n}^{-1}{ n
q_n}\big|S_m^*  \cap
 \big(\bigcup_{i =1}^{m-1} S_{i }^* \big)  \big|}  \nonumber \\ &
 =  e^{-n q_n} \sum_{r=0}^{X_n}
\mathbb{P}\bigg[\bigg|S_m\bcap \bigg(\bigcup_{i =1}^{m-1}S_{i
}^*\bigg)\bigg| = r \bigg] e^{{X_n}^{-1}{n q_nr} } . \label{SS}
\end{align}

For $r$ satisfying
\begin{align}
 0 &\leq r \leq |S_m|=X_n \nonumber
\end{align}
and
\begin{align}
 r & =  |S_m| + \bigg|\bigcup_{i =1}^{m-1}S_{i
}^*\bigg| - \bigg|S_m\bcup \bigg(\bigcup_{i =1}^{m-1}S_{i
}^*\bigg)\bigg|  \nonumber  \\
  &  \geq X_n + \bigg|\bigcup_{i =1}^{m-1}S_{i }^*\bigg|
- Y_n , \nonumber
\end{align}
as given in \cite[Eq. (36)]{mobihocQ1},
 we have
\begin{align}
 \mathbb{P}\bigg[\bigg|S_m\bcap \bigg(\bigcup_{i =1}^{m-1}S_{i
}^*\bigg)\bigg| = r \bigg]  &  \leq \frac{1}{r!} \bigg( \frac{m
{X_n}^2}{Y_n - X_n}\bigg)^r. \label{probsm2}
\end{align}


Applying (\ref{probsm2}) to (\ref{SS}), we establish
\begin{align}
&  \sum_{S_m^* } \mathbb{P}[ S_m = S_m^* ] e^{-{X_n}^{-1}nq_n{|S_m^*
\setminus \bigcup_{1\leq i \leq m-1} S_i^*|}} \nonumber \\ &  \leq
 e^{-n q_n} \sum_{r=0}^{X_n} \frac{1}{r!} \bigg( \frac{m {X_n}^2}{Y_n - X_n}\bigg)^r
\cdot e^{{X_n}^{-1}{n q_nr} }  \nonumber \\ &
 \leq  e^{-n q_n} \cdot e^{\frac{m {X_n}^2}{Y_n - X_n}
 \cdot e^{{X_n}^{-1}{n q_n}}} . \label{psnm}
\end{align}
From (\ref{eq_pe_lnnn})  and (\ref{PnKK}), it holds that
$\frac{{X_n}^2}{Y_n}\sim \frac{\ln n}{n}$, resulting in
\begin{align}
\frac{m {X_n}^2}{Y_n - X_n} &  \sim \frac{m{X_n}^2}{Y_n} \sim
\frac{m\ln n}{n}. \label{e1}
\end{align}
For an arbitrary $\epsilon > 0$, from (\ref{eq_pe_lnnn}), we obtain
$q_n \leq (1+\epsilon)\frac{\ln n}{n}$ for all $n$ sufficiently
large, which with condition $X_n \geq 2$ yields that for all $n$ sufficiently large,
\begin{align}
e^{{X_n}^{-1}{n q_n}} &  \leq e^{\frac{1}{2}(1+\epsilon)\ln n} =
n^{\frac{1}{2}(1+\epsilon)}.  \label{e2}
\end{align}
From (\ref{e1}) and (\ref{e2}), we get
\begin{align}
\frac{m {X_n}^2}{Y_n - X_n} \cdot e^{{X_n}^{-1}{n q_n}}  &  \leq
\frac{m\ln n}{n} \cdot [1+o(1)] \cdot n^{\frac{1}{2}(1+\epsilon)}
 \nonumber
\\ &  \leq m\ln n \cdot n^{\frac{1}{2}(\epsilon-1)}  \cdot [1+o(1)] .
\label{e3}
\end{align}
Since $\epsilon > 0$ is arbitrary, it follows from (\ref{e3}) that
for arbitrary $0<c<\frac{1}{2}$, then for all $n$ sufficiently
large, it is clear that
\begin{align}
\frac{m {X_n}^2}{Y_n - X_n} \cdot e^{{X_n}^{-1}{n q_n}} & \leq
n^{-c}. \label{e4}
\end{align}
Using (\ref{e4}) in (\ref{psnm}), for all $n$ sufficiently large, it follows that
\begin{align}
 \sum_{S_m^* } \mathbb{P}[ S_m = S_m^* ] e^{-{X_n}^{-1}nq_n{|S_m^*
\setminus \bigcup_{1\leq i \leq m-1} S_i^*|}} &  \leq e^{-n q_n}
\cdot e^{n^{-c}} .  \label{e5}
\end{align}
Substituting (\ref{e5}) into (\ref{HnmHnm1}), for all $n$
sufficiently large, we obtain
\begin{align}
 &\hspace{-2pt} Z_{m,n}  \nonumber  \\
 & \hspace{-3pt}\leq\hspace{-1.5pt} e^{-n q_n} \hspace{-1.5pt}\cdot\hspace{-1.5pt} e^{n^{-c}}
\hspace{-1.5pt}\cdot \hspace{-1.5pt}\sum_{\mathbb{S}_{m-1}^*} \hspace{-1pt}\mathbb{P}[\mathbb{S}_{m-1} \hspace{-1.5pt}=\hspace{-1.5pt}
\mathbb{S}_{m-1}^*]
e^{-{X_n}^{-1}nq_n{|\bigcup_{1\leq i \leq m-1} S_i^*|}} \nonumber \\
&  \hspace{-3pt}\leq\hspace{-1pt} e^{-n q_n} \hspace{-1pt}\cdot \hspace{-1pt}e^{n^{-c}} \hspace{-1pt}\cdot\hspace{-1pt} Z_{m-1,n}.
\end{align}
We then evaluate $Z_{2,n}$. By (\ref{ms}), it holds that
\begin{align}
& \hspace{-2pt}Z_{2,n}  \nonumber  \\
  &  \hspace{-3pt}=\hspace{-1.5pt}\sum_{S_1^*}\hspace{-1pt}\sum_{S_2^*} \hspace{-1pt}\Big\{ \hspace{-1pt}\mathbb{P}[(S_1  \hspace{-1pt}=\hspace{-1pt}
S_1^*)\hspace{-1pt}\bcap \hspace{-1pt}(S_2  \hspace{-1pt}=\hspace{-1pt} S_2^*)] \hspace{-1.5pt}\cdot\hspace{-1.5pt}e^{-{X_n}^{-1}nq_n{|S_1^* \bcup
S_2^*|}} \Big\}\nonumber \\  & = \sum_{S_1^*} \mathbb{P}[ S_1  =
S_1^* ] \sum_{S_2^*} \mathbb{P}[ S_2  = S_2^* ]
e^{-{X_n}^{-1}nq_n{|S_1^* \bcup S_2^*|}}. \label{mm1}
\end{align}
Setting $m=2$ in (\ref{e5}), for all $n$ sufficiently large, we derive
\begin{align}
\sum_{S_2^*} \mathbb{P}[ S_2  = S_2^* ] e^{-{X_n}^{-1}nq_n{| S_2^*
\setminus S_1^*|}} & \leq e^{-n q_n} \cdot e^{n^{-c}} . \nonumber
\end{align}
Then for all $n$ sufficiently large, it follows that
\begin{align}
& \sum_{S_2^*} \mathbb{P}[ S_2  = S_2^* ] e^{-{X_n}^{-1}nq_n{|S_1^*
\bcup S_2^*|}}  \nonumber  \\
  &  =  e^{-n q_n} \sum_{S_2^*} \mathbb{P}[ S_2  =
S_2^* ] e^{-{X_n}^{-1}nq_n{| S_2^* \setminus S_1^*|}}  \nonumber \\
& \leq  e^{-2n q_n} \cdot e^{n^{-c}} . \label{e6}
\end{align}

From (\ref{mm1}) and (\ref{e6}), for all $n$ sufficiently large, we
obtain
\begin{align}
  Z_{m,n} & \leq \big(e^{-n q_n} \cdot e^{n^{-c}}\big)^{m-2}
\cdot Z_{2,n} \nonumber \\
&  \leq \big(e^{-n q_n} \cdot e^{n^{-c}}\big)^{m-2} \cdot e^{-2n
q_n} \cdot e^{n^{-c}}
\nonumber \\
&  \leq e^{-mn q_n} \cdot e^{(m-1)n^{-c}} .
\end{align}
Letting $n \to \infty$, we finally establish
\begin{align}
  Z_{m,n} & \leq e^{-m n q_n} \cdot
 [1+o(1)] ; \nonumber
\end{align}
i.e., (\ref{ms}) is proved. Then as explained above, (\ref{sstar})
holds; and then (\ref{prop2}) follows.

\subsection{The Proof of (\ref{prop1})}

Again let $w$ be an arbitrary node in $\overline{V_m}$. We have
\begin{align}
& \mathbb{P}\bigg[\bigg(\bigcap_{1\leq i <j \leq m}
(N_{ij}=\emptyset)\bigg)  \bcap \bigg(\bigcap_{1\leq i \leq
m}(|N_{i}| = h)\bigg) \bgiven \mathbb{S}_m= \mathbb{S}_m^*\bigg]
 \label{h} \\ & = \frac{(n-m)!}{(h!)^m(n-m-hm)!}  \nonumber \\
  & \quad \times \prod_{1\leq i \leq m}\left(\left\{\mathbb{P}
  \left[\begin{array}{l}w\leftrightarrow v_i,\\\text{but }w\leftrightarrow\text{none of}\\\text{nodes
in }V_m \setminus \{v_i\}\end{array}\Bigg|\hspace{3pt} \mathbb{S}_m= \mathbb{S}_m^*\right]\right\}^h\right)  \nonumber \\
& \quad \times \big\{\mathbb{P}[w\leftrightarrow \text{none of
nodes in }V_m \given \mathbb{S}_m= \mathbb{S}_m^*]\big\}^{n-m-hm}
\label{y}
\end{align}
and
\begin{align}
P_a  & = \sum_{\mathbb{S}_m^*:\hspace{2pt}\bigcap_{1\leq i <j \leq
m} (S_{ij}^*=\emptyset)} \Big\{ \mathbb{P}[\mathbb{S}_m =
\mathbb{S}_m^*] \cdot (\ref{h})
 \Big\} \label{pra} ,
\end{align}
where $S_{ij}^{*} : = S_{i}^{*} \cap S_{j}^{*}$.

For $i=1,2,\ldots,m$, under
$\mathbb{S}_m^*:\hspace{2pt}\bigcap_{1\leq i <j \leq m}
(S_{ij}^*=\emptyset)$, we have
\begin{align}
& \mathbb{P}
  [w\leftrightarrow v_i,\text{ but none of nodes
in }V_m \setminus \{v_i\}\given \mathbb{S}_m= \mathbb{S}_m^*]  \nonumber \\
& \geq \mathbb{P}
  [w\leftrightarrow v_i \given \mathbb{S}_m= \mathbb{S}_m^*]
 \nonumber \\
&  \quad - \sum_{\begin{subarray}{c} 1 \leq j \leq m \\ j\neq
i\end{subarray}} \mathbb{P}
  [w\leftrightarrow \text{both }v_i\text{ and }v_j
  \given \mathbb{S}_m= \mathbb{S}_m^*] \nonumber \\
&\geq q_n - (m-1) \cdot 2{q_n} ^2 \quad \text{(by \cite[Lemma
3]{mobihocQ1})} . \label{ps2}
\end{align}
Substituting (\ref{sstar3}) and (\ref{ps2}) to (\ref{y}), and then
from (\ref{pra}), we obtain
\begin{align}
P_a  & \geq \frac{(n-m-hm)^{hm}}{(h!)^m} \cdot [q_n - 2(m-1) q_n
^2]^{hm}    \nonumber \\
&  \quad   \times (1-mq_n)^{n-m-hm} \sum_{\mathbb{S}_m^*:\hspace{2pt}\bigcap_{1\leq i
<j \leq m} (S_{ij}^*=\emptyset)}  \mathbb{P}[\mathbb{S}_m =
\mathbb{S}_m^*]. \nonumber
\end{align}
Then from (\ref{eq_pe_lnnn}), it further hold that
\begin{align}
P_a  & \geq \frac{n^{hm}}{(h!)^m} \cdot (q_n)^{hm}
  \cdot  e^{- m n q_n}  \nonumber \\
&  \quad \times [1-o(1)] \cdot
  \mathbb{P}\bigg[\bigcap_{1\leq i <j \leq m} (S_{ij}=\emptyset)\bigg]
    . \label{poa1}
\end{align}

From (\ref{sstar4}), under
$\mathbb{S}_m^*:\hspace{2pt}\bigcap_{1\leq i <j \leq m}
(S_{ij}^*=\emptyset)$, it holds that
\begin{align}
 \mathbb{P}[w\leftrightarrow \text{none of nodes in }V_m \given
\mathbb{S}_m= \mathbb{S}_m^*] & \leq e^{- m q_n} . \label{ps3a}
\end{align}

For each $i=1,2,\ldots,m$, we have
\begin{align}
& \mathbb{P}
  [w\leftrightarrow v_i,\text{ but }w\leftrightarrow \text{none of nodes
in }V_m \setminus \{v_i\}\given \mathbb{S}_m= \mathbb{S}_m^*]  \nonumber \\
& \leq \mathbb{P}
  [w\leftrightarrow v_i \given \mathbb{S}_m= \mathbb{S}_m^*] =
  q_n. \label{ps}
\end{align}
Substituting (\ref{ps}) and (\ref{ps3a}) to (\ref{y}), and then from
(\ref{pra}), we obtain 
\begin{align}
P_a  & \hspace{-1.5pt}\leq\hspace{-1.5pt} \frac{n^{hm}}{(h!)^m} \hspace{-1.5pt}\cdot\hspace{-1.5pt} (q_n)^{hm}
  \hspace{-1.5pt}\cdot\hspace{-1.5pt}  e^{- m n q_n} \hspace{-1.5pt}\cdot\hspace{-1.5pt} \sum_{\mathbb{S}_m^*:\hspace{2pt}\hspace{-1.5pt}\bigcap_{1\leq i
<j \leq m} (S_{ij}^*=\emptyset)}  \hspace{-1.5pt}\mathbb{P}[\mathbb{S}_m \hspace{-1.5pt}=\hspace{-1.5pt}
\mathbb{S}_m^*] \nonumber \\
&   =  \frac{n^{hm}}{(h!)^m} \cdot (q_n)^{hm}
  \cdot  e^{- m n q_n} \cdot
  \mathbb{P}\bigg[\bigcap_{1\leq i <j \leq m} (S_{ij}=\emptyset)\bigg]
 .  \label{poa2}
\end{align}

From (\ref{poa1}) and (\ref{poa2}),  we obtain\vspace{-1pt}
\begin{align}
P_a  &   \sim \frac{n^{hm}}{(h!)^m} \cdot (q_n)^{hm}
  \cdot  e^{- m n q_n} \cdot
  \mathbb{P}\bigg[\bigcap_{1\leq i <j \leq m} (S_{ij}=\emptyset)\bigg].
  \label{y2} \vspace{-1pt}
\end{align}
By the union bound, it is clear that\vspace{-1pt}
\begin{align}
& \mathbb{P}\bigg[\bigcap_{1\leq i <j \leq m} (S_{ij}=\emptyset)\bigg]\vspace{-1pt}
\nonumber \\ & \quad = 1 - \mathbb{P}\bigg[\bigcup_{1\leq i <j \leq m} (S_{ij}\neq
\emptyset)\bigg]\vspace{-1pt} \nonumber \\  &  \quad \geq 1 - \sum_{1\leq i <j \leq m}
\mathbb{P}[S_{ij}\neq \emptyset] = 1 - \binom{m}{2}q_n. \label{m2ps}\vspace{-1pt}
\end{align}
From (\ref{eq_pe_lnnn}) and (\ref{m2ps}), since a
probability is at most $1$, we get\vspace{-1pt}
\begin{align}
\lim_{n \to \infty}\mathbb{P}\bigg[\bigcap_{1\leq i <j \leq m}
(S_{ij}=\emptyset)\bigg] & = 1 .\vspace{-1pt} \label{m2ps2}
\end{align}
Using (\ref{m2ps2}) in (\ref{y2}), we establish (\ref{prop1}).

\section{Numerical Experiments} \label{sec:expe}

\begin{figure}[!t]
  \centering
 \includegraphics[width=0.45\textwidth]{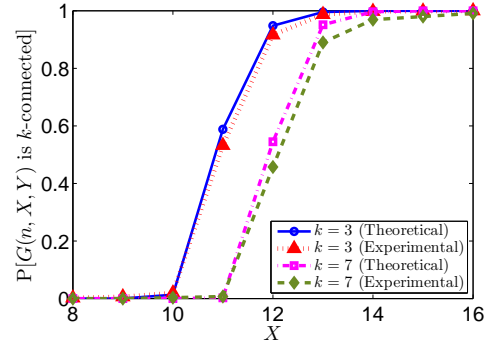}
\caption{A plot for the probability of $k$-connectivity in graph
$G(n,X,Y)$ with $k = 3,7$ under $n=3,000$ and
$Y=20,000$.\vspace{-5pt}
         }
\label{fig}
\end{figure}

We present numerical experiments to back up our
theoretical results. %
%
%
%
 Figure \ref{fig} depicts the probability that graph
$G(n,X,Y)$ is $k$-connected. We let $X$ vary, with other parameters
fixed at $n=3,000$, $Y=30,000$ and $k=3 ,7$. The empirical
probabilities corresponding to the experimental curves are obtained
as follows: we count the times of $k$-connectivity out of $500$
independent samples of $G(n,X,Y)$, and derive the empirical
probability through dividing its corresponding count by $500$. For
the theoretical curves, we first compute $\alpha$ by setting \vspace{1pt} the edge probability $1-
\frac{\binom{Y- X}{X}}{\binom{Y}{X}}$ (viz., (\ref{qnp}) in the Appendix) \vspace{-1pt}
 as $ \frac{\ln  n + {(k-1)} \ln \ln n + {\alpha}}{n}$ and then use
 $e^{- \frac{e^{-\alpha}}{(k-1)!}}$ as the theoretical
 value for the probability of $k$-connectivity. Figure \ref{fig}
 confirms our analytical results as the experimental and theoretical
 curves are close.

\section{Related Work}  \label{related}

For a random key graph $G(n,X_n,Y_n)$, Rybarczyk \cite{ryb3} derives
the asymptotically exact probabilities of connectivity and of the
property that the minimum node degree is no less than $1$, covering
a weaker form of the results -- the zero-one laws which are also
 obtained in \cite{r1,yagan}. As demonstrated in \cite{ryb3}, in
$G(n,X_n,Y_n)$ with $X_n \geq 2$, $\frac{{X_n}^2}{Y_n}=\frac{\ln n +
{\alpha_n}}{n}$ and $\lim_{n \to \infty} \alpha_n = \alpha ^* $, the
probability of connectivity and that of the minimum degree being at
least $1$ both approach to $e^{- e^{-\alpha ^*}}$ as $n\to \infty$.
Rybarczyk \cite{zz} implicitly obtains zero-one laws (but not the
asymptotically exact probabilities) for $k$-connectivity and for the
property that the minimum degree is at least $k$. The implicit
result is that if $X_n \Theta(n
^{\beta})$ for some $\beta
>0$ and $\frac{{X_n}^2}{Y_n}=\frac{\ln n + (k-1) \ln \ln n +
{\alpha_n}}{n}$, graph $G(n,X_n,Y_n)$ has (resp., does not have) the
two properties with probability approaching to $1$, given that $
{\alpha_n} $ tends to $\infty$ (resp., $-\infty$) as $n\to \infty$.
Our Corollary \ref{COR_RKGk} significantly improves her result
\cite{zz} in the following two aspects: (i) we cover the wide range of $ X_n \geq 2$
all $n$ sufficiently large, instead of the much stronger condition
$X_n = \Omega\big((\ln n)^3\big)$ in \cite{zz} (note that the analysis under $X_n=1$ is trivial), and
(ii) we establish not only zero--one laws for $k$-connectivity
  and the minimum degree, but also the asymptotically exact
  probabilities. The latter results are not given by Rybarczyk \cite{zz}. Recently, we \cite{JZCDC} give
  the asymptotically exact
  probability of $k$-connectivity in graph $G(n,X_n,Y_n)$ under $X_n = \Omega(\sqrt{\ln n}\hspace{1.5pt})$ through a rather involved proof. We improve this result
  to cover $ X_n \geq 2$ through a simpler proof and fill the gap where $X_n$ is at least $2$, but is not $ \Omega(\sqrt{\ln n}\hspace{1.5pt})$. This improvement is of technical interest as well as of practical importance since random key graphs have been used in diverse applications including modeling the Eschenauer--Gligor random key predistribution scheme (the most recognized solution to secure communication in wireless sensor networks). 
  

%
%

\section{Conclusion}
\label{sec:Conclusion}

In this paper, for a random key graph, we derive the asymptotically
exact probabilities for two properties with an arbitrary $k$: (i)
the graph is $k$-connected; and (ii) each node has at least $k$
neighboring nodes. Numerical experiments are in accordance with our
analytical results.

 \renewcommand\baselinestretch{.95}

\small

\bibliographystyle{abbrv}
\bibliography{related}

%

%

%
%
%
%
%
%
%

\normalsize

 \renewcommand\baselinestretch{1}

\appendix

\subsection{Establishing Corollary \ref{COR_RKGk} given Theorem
\ref{THM_RKGk}:}

As noted in Remark \ref{rem1}, we can use an extra condition
$\beta_n = \pm O(\ln \ln n)= \pm o(\ln n)$ in establishing Corollary \ref{COR_RKGk}.

With $q_n$ denoting the probability that there exists an edge
between two nodes in graph $G(n, X_n, Y_n)$, as shown in previous
work \cite{r1,ryb3,yagan}, we have
\begin{align}
q_n &
 = 1- \frac{\binom{Y_n- X_n}{X_n}}{\binom{Y_n}{X_n}}. \label{qnp}
\end{align}
Further, it holds from \cite[Lemma 8]{ZhaoYaganGligor} that
\begin{align}
q_n & = \frac{{X_n}^2}{Y_n} \pm O\Bigg(\bigg( \frac{{X_n}^2}{Y_n}
\bigg)^2 \Bigg). \label{PnKK}
\end{align}
From (\ref{thm_eq_psXnPn}) and $\beta_n = \pm o(\ln n)$, it follows
that
\begin{align}
\frac{{X_n}^2}{Y_n} & =  O\bigg(\frac{\ln  n}{n}\bigg).
\label{thm_eq_psXnPn2}
\end{align}
Substituting (\ref{thm_eq_psXnPn}) and (\ref{thm_eq_psXnPn2}) to
(\ref{PnKK}), we further obtain
\begin{align}
q_n & = \frac{\ln  n + {(k-1)} \ln \ln n + {\beta_n}}{n} \pm
O\Bigg(\bigg(\frac{\ln n}{n}\bigg)^2 \Bigg). \nonumber \\
&   = \frac{\ln  n + {(k-1)} \ln \ln n + {\beta_n} \pm
O\big(n^{-1}(\ln n)^2\big)}{n}. \label{thm_eq_psXnPn3}
\end{align}
With $\alpha_n$ defined by (\ref{thm_eq_ps}), from (\ref{thm_eq_ps})
and (\ref{thm_eq_psXnPn3}), it holds that
\begin{align}
\alpha_n & = {\beta_n} \pm O\big(n^{-1}(\ln n)^2\big). \nonumber
\end{align}
Therefore, $\lim_{n \to \infty} \alpha_n$ exists if and only if $\lim_{n \to
\infty} {\beta_n}$ exists, and
 $\lim_{n \to \infty} \alpha_n = \lim_{n \to
\infty} {\beta_n}$ holds. Then
 Theorem \ref{THM_RKGk} clearly implies Corollary
\ref{COR_RKGk}. \hfill $\blacksquare$

\subsection{Lemma 4 to confine $|\alpha_n|$ as $O(\ln \ln n)$ in Theorem 1}

\begin{lem} \label{lem-only-prove-lnlnn-1}

For a random key graph
$G(n,X_n,Y_n)$ under $X_n \geq 2$ and $q_n =  \frac{\ln  n + {(k-1)} \ln \ln n + {\widetilde{\alpha_n}}}{n}$, the following results hold:

(i) If $\lim_{n \to \infty}\alpha_n = -\infty$, there exists graph $G(n,\widetilde{X_n},\widetilde{Y_n})$ under $\widetilde{X_n} \geq 2$ and
$\widetilde{q_n} =  \frac{\ln  n + {(k-1)} \ln \ln n + {\widetilde{\alpha_n}}}{n}$ with $\lim_{n \to \infty}\widetilde{\alpha_n} = -\infty$ and $\widetilde{\alpha_n} = -O(\ln \ln n)$ (\hspace{1pt}$\widetilde{q_n} $ is the edge probability of $G(n,\widetilde{X_n},\widetilde{Y_n})$),
such that there exists a graph coupling\footnote{As used
by Rybarczyk \cite{zz,2013arXiv1301.0466R}, a coupling of two random graphs $G_1$ and
$G_2$ means a probability space on which random graphs $G_1'$ and
$G_2'$ are defined such that $G_1'$ and $G_2'$ have the same
distributions as $G_1$ and $G_2$, respectively. If $G_1'$ is a spanning subgraph
(resp., spanning supergraph) of $G_2'$, we say that under the coupling, $G_1$ is a spanning subgraph
(resp.,  spanning supergraph) of $G_2$, which yields that for any monotone increasing property $\mathcal {I}$, the probability of $G_1$ having $\mathcal {I}$ is at most (reap., at least) the probability of $G_2$ having $\mathcal {I}$.} under which
$G(n,X_n,Y_n)$ is a spanning subgraph of $G(n,\widetilde{X_n},\widetilde{Y_n})$.

(ii) If $\lim_{n \to \infty}\alpha_n = \infty$, there exists graph $G(n,\widehat{X_n},\widehat{Y_n})$ under $\widehat{X_n} \geq 2$ and 
$\widehat{q_n}   = \frac{\ln  n + {(k-1)} \ln \ln n + {\widehat{\alpha_n}}}{n}$
with $\lim_{n \to \infty}\widehat{\alpha_n} = \infty$ and $\widehat{\alpha_n} = O(\ln \ln n)$ (\hspace{1pt}$\widehat{q_n} $ is the edge probability of $G(n,\widehat{X_n},\widehat{Y_n})$),
such that there exists a graph coupling under which
$G(n,X_n,Y_n)$ is a spanning supergraph of $G(n,\widehat{X_n},\widehat{Y_n})$.

\end{lem}

\subsection{Lemma 5 to confine $|\beta_n|$ as $O(\ln \ln n)$ in Corollary 1}

\begin{lem} \label{lem-only-prove-lnlnn-2}

For a random key graph
$G(n,X_n,Y_n)$ under $X_n \geq 2$ and $\frac{{\widetilde{X_n}}^2}{\widetilde{Y_n}} =  \frac{\ln  n + {(k-1)} \ln \ln n + {\widetilde{\beta_n}}}{n}$, the following results hold:

(i) If $\lim_{n \to \infty}\beta_n = -\infty$, there exists graph $G(n,\widetilde{X_n},\widetilde{Y_n})$ under $\widetilde{X_n} \geq 2$ and
$\frac{{\widetilde{X_n}}^2}{\widetilde{Y_n}} =  \frac{\ln  n + {(k-1)} \ln \ln n + {\widetilde{\beta_n}}}{n}$ with $\lim_{n \to \infty}\widetilde{\beta_n} = -\infty$ and $\widetilde{\beta_n} = -O(\ln \ln n)$,
such that there exists a graph coupling under which
$G(n,X_n,Y_n)$ is a spanning subgraph of $G(n,\widetilde{X_n},\widetilde{Y_n})$.

(ii) If $\lim_{n \to \infty}\beta_n = \infty$, there exists graph $G(n,\widehat{X_n},\widehat{Y_n})$ under $\widehat{X_n} \geq 2$ and 
$  \frac{{\widehat{X_n}}^{2}}{{\widehat{Y_n}}}    = \frac{\ln  n + {(k-1)} \ln \ln n + {\widehat{\beta_n}}}{n}$
with $\lim_{n \to \infty}\widehat{\beta_n} = \infty$ and $\widehat{\beta_n} = O(\ln \ln n)$,
such that there exists a graph coupling under which
$G(n,X_n,Y_n)$ is a spanning supergraph of $G(n,\widehat{X_n},\widehat{Y_n})$.

\end{lem}


\end{document}